\documentclass[twocolumn]{emulateapj-rtx4}
\usepackage{graphicx}
\usepackage{subfigure}
\usepackage{amsmath}
\usepackage{amsfonts}

\newcommand{\cgal}{\texttt{CGAL}\ }
\newcommand{\mm}[1]      {\ifmmode{#1}\else{\mbox{\(#1\)}}\fi}
\newcommand{\Manifold}   {\mm{{\mathbb M}}}

\newcommand{\Betti}      {\mm{{\beta}}}
\newcommand{\Euler}      {\mm{\chi}}
\newcommand{\genus}      {\mm{G}}
\newcommand{\genusalt}   {\mm{g}}
\newcommand{\diff}       {\mm{\rm \,d}}

\usepackage[usenames]{color}

\definecolor{MyGrey10}{rgb}{0.10,0.10,0.10}
\definecolor{MyGrey30}{rgb}{0.30,0.30,0.30}
\definecolor{MyGrey80}{rgb}{0.80,0.80,0.80}
\definecolor{MyGrey90}{rgb}{0.90,0.90,0.90}
\definecolor{MyGrey95}{rgb}{0.95,0.95,0.95}
\definecolor{MyGrey98}{rgb}{0.98,0.98,0.98}
\definecolor{MyGrey100}{rgb}{1.0,1.0,1.0}
\definecolor{MyYellow}{rgb}{1.0,1.0,0.0}

\begin{document}
\title{Probing Dark Energy with Alpha Shapes and Betti Numbers}
\author{Rien~van~de~Weygaert\altaffilmark{1}, Pratyush Pranav\altaffilmark{1}, Bernard J.T. Jones\altaffilmark{1}, E.G.P. (Patrick) Bos\altaffilmark{1}, Gert Vegter\altaffilmark{2}, \\ Herbert Edelsbrunner\altaffilmark{3}\altaffilmark{,4}\altaffilmark{,5}, Monique Teillaud\altaffilmark{6},Wojciech A. Hellwing\altaffilmark{7}\altaffilmark{,8}, Changbom Park\altaffilmark{9}, \\  
Johan Hidding\altaffilmark{1} and Mathijs Wintraecken\altaffilmark{2}}

\altaffiltext{1}{Kapteyn Astron. Inst., University of Groningen, PO Box 800, 9700 AV Groningen, The Netherlands; weygaert@astro.rug.nl}
\altaffiltext{2}{Johann Bernoulli Institute for Mathematics and Computer Science, University of Groningen,\\ P.O. Box 407, 9700 AK Groningen, the Netherlands}
\altaffiltext{3}{IST Austria, Am Campus 1, 3400 Klosterneuburg, Austria}
\altaffiltext{4}{Duke University, Computer Science Department, Box 90129, Durham, NC 27708, USA} 
\altaffiltext{5}{Geomagic Inc., Research Triangle Park,  NC 27709, USA}
\altaffiltext{6}{INRIA Sophia Antipolis-M\'editerran\'ee, route des Lucioles, BP 93, 06902 Sophia Antipolis Cedex, 
France}
\altaffiltext{7}{Institute of Computational Cosmology, Department of Physics, Durham University, South Road, Durham DH1 3LE, United Kingdom}
\altaffiltext{8}{Interdisciplinary Centre for Mathematical and Computational Modeling, University of Warsaw,\\ ul. Pawinskiego 5a, 
02-106 Warsaw, Poland}
\altaffiltext{9}{School of Physics, Korea Institute for Advanced Study, Seoul 130-722, Korea}

\begin{abstract}
We introduce a new descriptor of the weblike pattern in the distribution of galaxies and matter: the scale dependent 
Betti numbers which formalize the topological information content of the cosmic mass distribution.
While the Betti numbers do not fully quantify topology, they extend the information beyond conventional cosmological 
studies of topology in terms of genus and Euler characteristic used in earlier analyses of cosmological models. The richer 
information content of Betti numbers goes along with the availability of fast algorithms to compute them.  When measured as a 
function of scale they provide a ``Betti signature'' for a point distribution that is a sensitive yet robust discriminator 
of structure.  The signature is highly effective in revealing differences in structure arising in different cosmological 
models, and is exploited towards distinguishing between different dark energy models and may likewise be used to 
trace primordial non-Gaussianities. 

In this study we demonstrate the potential of Betti numbers by studying their behaviour in simulations of cosmologies 
differing in the nature of their dark energy.  
\end{abstract}


\maketitle

\section{Introduction}
The large scale distribution of matter mapped by galaxy surveys like the 2dF, SDSS and 
2MASS redshift surveys \citep{colless2003,gott2005,huchra2005} shows a complex 
network of interconnected filamentary galaxy associations. This network, which has become known as 
the {\it Cosmic Web} \citep{bondweb1996}, contains structures from a few megaparsecs up to tens and 
even hundreds of megaparsecs in size. Galaxies and mass exist in a wispy web-like spatial arrangement 
consisting of dense compact clusters, elongated filaments, and sheet-like walls, amidst large near-empty 
voids, with similar patterns existing at earlier epochs, albeit over smaller scales 
\citep[for a review, see][]{weybond2008}.

Given the structural richness of the Cosmic Web, in this paper we address the question of how we can 
exploit the information content in the structure of the Cosmic Web to differentiate cosmological models having 
different dark energy content. Different dark energy models, viewed at the same redshift, show subtly different 
structures in the matter distribution. This is simply a consequence of the different growth rates 
for cosmic structure in the different models.  

Despite the multitude of descriptions, it has remained a major challenge to characterize the structure, geometry and 
topology of the Cosmic Web. Many of the attempts to describe, let alone identify, the features 
and components of the Cosmic Web have been of a rather heuristic nature. Moreover, the overwhelming complexity of both the 
individual structures as well as their connectivity, the lack of structural symmetries, its intrinsic multi-scale 
nature and the wide range of densities in the cosmic matter distribution has prevented the use of simple and 
straightforward methods. 

\begin{figure}[h]
  \begin{center}
  \includegraphics[width=3.25in]{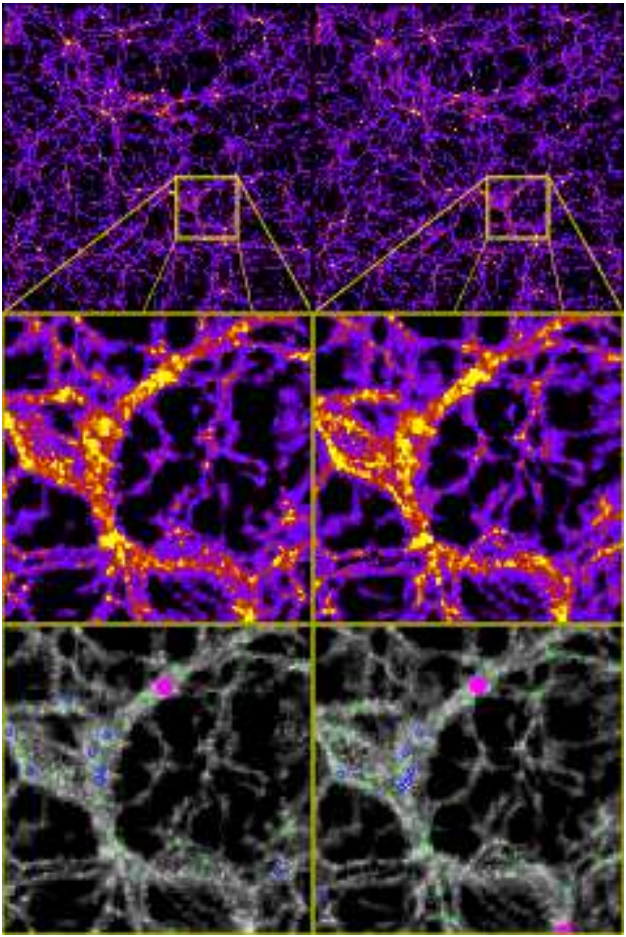}  
  \caption{Slices of thickness $0.2h^{-1} \mathrm{Mpc}$ from two cosmological models: $\Lambda$CDM (left), SUGRA (right). 
Top: $200h^{-1} \times 200h^{-1} \mathrm{Mpc}$ slices. Middle \& Bottom: zoom-in, $40h^{-1}\times 40h^{-1} \mathrm{Mpc}$ 
slices. Middle: density field. Bottom: halo distribution, superimposed. 
}
  \end{center}
  \label{fig:SIMULATIONS}
\end{figure}
Here we introduce a new topological measure that is particularly suited to differentiating web-like 
structures. 
To this end, we advance the topological characterization of the Cosmic Web to a more 
complete description, the homology of the distribution as measured by the scale-dependent Betti numbers 
of the sample \citep{edelsbrunner1994,edelsbrunner2010,robins2006,weygaert2010,weygaert2011}. 
The Betti numbers underlie genus analysis and may be regarded as fundamental topological 
structure indicators that specifically characterise the homology of the distribution. 
While a full quantitative characterization of the  topology of the cosmic mass distribution may not 
be feasible, the homology is an attractive compromise, providing a usefully detailed summary measurement 
of topology with relatively low computational overhead \citep[e.g.][]{delfinado1993,edelsbrunner1994}.

\subsection{Dark energy and the Cosmic Web}
The parameters for what might now be called ``the standard model for cosmology'' have been established 
with remarkable precision.  However, there remains the great mystery of the nature of the so-called ``dark energy'' (DE) 
which appears to make up some 73\% of the total cosmic energy density.  The simplest model for the dark energy is 
Einstein's cosmological constant, $\Lambda$: this makes a constant, time independent, contribution to the total energy 
density in the Friedman-Lemaitre equations.   Models based on this are referred to as ``$\Lambda$CDM models''.  
However, there are numerous, possibly more plausible, models in which the dark energy evolves as a function of time.  
These models are generally described in terms of a time or redshift dependent function $w(z)$ that describes the history 
of equation of state of the dark energy component:
\begin{equation}
p = w(z) \rho c^2
\end{equation}
Different models for dark energy produce different functions $w(z)$. 
The Einstein cosmological constant corresponds to $w(z) = -1$. 

The aim of this paper is to compare,  at redshift $z=0$, the topology of the cosmic structure in simulations of three cosmological 
models having different dark energy content that evolved from the same initial conditions (fig.~\ref{fig:SIMULATIONS}). The 
equation of state of the dark energy models are those of the standard $\Lambda$CDM model and two quintessence models. The latter 
assume that the Universe contains an evolving quintessence scalar field, whose energy content manifests itself as dark energy. 
The two quintessence models are the \emph{Ratra-Peebles (RP) model} and a \emph{SUGRA model} \citep[for a detailed description, 
see][]{ratra1988,amendola2000,brax1999,perrotta2000,deboni11,bos2011}.

\section{Topology and Homology}
\subsection{Betti Numbers}
Formally, homology groups and Betti numbers characterize the topology of a space in terms of the relationship between the 
cycles and boundaries we find in the space\footnote{Assuming the space is given as a simplicial complex,
a \emph{$p$-cycle} is a $p$-chain with empty boundary, where a \emph{$p$-chain}, $\gamma$, is a sum of $p$-simplices.
The standard notation is $\gamma = \sum a_i \sigma_i$, where the $\sigma_i$ are the $p$-simplices and the $a_i$ are 
the {\it coefficients}. For example, a 1-cycle is a closed loop of edges, or a finite union of such loops,
and a 2-cycle is a closed surface, or a finite union of such surfaces. Adding two $p$-cycles, we get another $p$-cycle, 
and similar for the $p$-boundaries. Hence, we have a group of $p$-cycles and a group of $p$-boundaries.}\citep{munkres1995,
zomorodian2005,edelsbrunner2010}. 

For example, if the space is a $d$-dimensional manifold, $\Manifold$, we have cycles and boundaries of dimension $p$ from $0$ to $d$.
Correspondingly, $\Manifold$ has one homology group $H_p (\Manifold)$ for each of $d+1$ dimensions, $0 \leq p \leq d$. By taking 
into account that two cycles should be considered identical if they differ by a boundary, one ends up with a group $H_p (\Manifold)$
whose elements are the equivalence classes of $p$-cycles.
The \emph{rank} of the homology group $H_p (\Manifold)$ is the $p$-th \emph{Betti number},
$\Betti_p = \Betti_p (\Manifold)$.

In heuristic - and practical - terms, the Betti numbers count topological features
and can be considered as the number of $p$-dimensional holes. When 
talking about a surface in $3$-dimensional space, the zeroth Betti number, $\Betti_0$, counts the components or ``islands'', 
the first Betti number, $\Betti_1$, counts the tunnels,
and the second Betti number, $\Betti_2$, counts the enclosed voids. All other Betti numbers are zero.

\begin{figure*}[t]
  \begin{center}
 \includegraphics[width=0.85\textwidth]{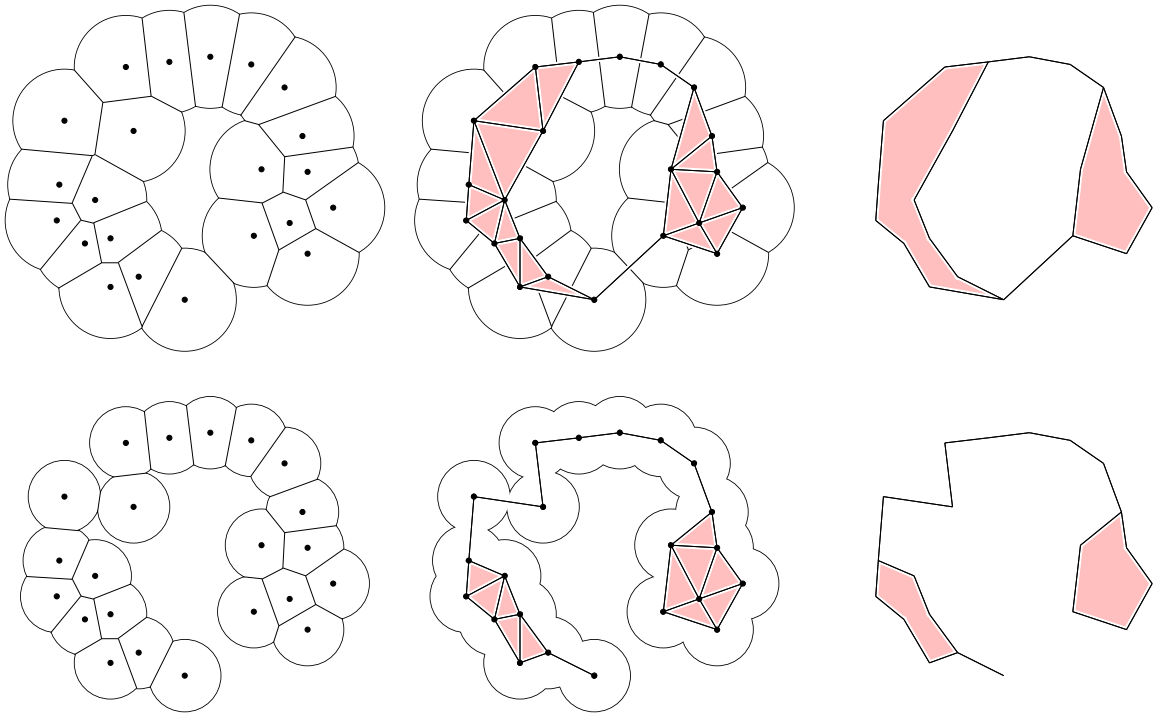}  
  \caption{Illustration of alpha shapes.  For two different values of $\alpha$ (top: large $\alpha$; bottom: small $\alpha$), 
    we show the relation between the $2$-dimensional point distribution, the value of $\alpha$,
    and the resulting alpha shape.  Around each point in the sample, we draw a circle of radius $\alpha$.
    The outline of the corresponding Voronoi tessellation is indicated by the edges (left).
    All Delaunay simplices \emph{dual} to the decomposition of the union of disks by the Voronoi
    polygons are shown in black (vertices and edges) and red (triangles) (center).  The resulting alpha shape 
    is shown on the right.}
  \end{center}
  \label{fig:ALPHASHAPE}
\end{figure*}

\subsection{Genus and the Euler characteristic} 
Numerous cosmological studies have considered the \emph{genus} of the isodensity surfaces
defined by the megaparsec galaxy distribution \citep{gott1986,hamilton1986,choi2010},
which specifies the number of handles defining the surface\footnote{For consistency,
   it is important to note that the definition in previous 
   cosmological topology studies slightly differs from this.
   The genus, $\genusalt$, in these studies has been defined as the number of holes minus the 
   number of connected regions: $\genusalt=\genus-c$.}. 
The genus has a simple relation to the Euler characteristic, $\Euler$, of the isodensity surface. 
Consider a $3$-manifold subset $\Manifold$ of the Universe and its boundary, $\partial \Manifold$, 
which is a $2$-manifold. With $\partial \Manifold$ consisting of $c=\Betti_0 (\partial \Manifold)$ components,
the Gauss-Bonnet Theorem states that the genus of the surface is given by
\begin{eqnarray}
  \genus  &=&  c - \frac{1}{2} \Euler (\partial \Manifold) ,
\end{eqnarray}
where the Euler characteristic $\Euler (\partial \Manifold)$ is
the integrated Gaussian curvature of the surface 
\begin{eqnarray}
  \Euler (\partial \Manifold)  &=&  \frac{1}{2 \pi} \oint_x \frac{\diff x}{R_1 (x) R_2 (x)} . 
  \label{eq:euler}
\end{eqnarray}
Here $R_1 (x)$ and $R_2 (x)$ are the principal radii of curvature at point $x$ of the surface.
The integral of the Gaussian curvature is invariant under continuous deformation of the surface:
perhaps one of the most surprising results in differential geometry.

According to the Euler-Poincar\'{e} Formula, the Euler characteristic of the manifold $\Manifold$ is 
equal to the alternating sum of its Betti numbers, 
\begin{equation}
  \Euler (\Manifold)=\Betti_0(\Manifold)-\Betti_1(\Manifold)+\Betti_2 (\Manifold)\,.
\end{equation}
For practical circumstances the third Betti number vanishes, $\Betti_3 (\Manifold)=0$. 
The Euler characteristic of the boundary, $\partial \Manifold$, is directly proportional to that 
of the Euler characteristic of the 3-manifold, $\Manifold$, \citep[][p.223]{seifert1934}
 \begin{eqnarray}
  \Euler (\partial \Manifold)  
                               &=&  2 \Euler (\Manifold) .
\label{eq:betti2-3}
\end{eqnarray}
The combination of these equations for the Euler characteristic establishes a fundamental 
relationship between differential geometry and algebraic topology. 

In the analysis described in this paper, we restrict ourselves to the three Betti numbers, 
$\Betti_0$, $\Betti_1$, and $\Betti_2$, of the $3$-manifolds with boundary defined by the cosmic 
mass distribution. In an upcoming paper, we study genus and Betti number characteristics of 
Gaussian fields \citep{park2011}.

\section{Alpha Shapes}
\label{sec:alphashape}
One of the key concepts in the field of Computational Topology is \emph{alpha shapes},
introduced by Edelsbrunner and collaborators \citep{edelsbrunner1983,mueckephd1993,edelsbrunner1994,
edelsbrunner2009}. They generalize the convex hull of a point set and are concrete geometric objects
that are uniquely defined for a particular point set $S$ and a scale $\alpha$.
For their definition, we look at the union of balls of radius $\alpha$ centered on 
the points in the set, and its decomposition by the corresponding Voronoi tessellation.
The \emph{alpha complex} consists of all Delaunay simplices that record
the subsets of Voronoi cells that have a non-empty common intersection
within this union of balls (fig.~\ref{fig:ALPHASHAPE}).

The \emph{alpha complex} thus consists of all simplices in the Delaunay triangulation that have an empty 
circumsphere with radius less than or equal to $\alpha$. Here ``empty'' means that the open ball bounded by 
the sphere does not include any points of $S$. For the extreme value $\alpha=0$, the alpha complex merely 
consists of the vertices of the point set. There is also a smallest value, $\alpha_{\max}$, such that for 
$\alpha \geq \alpha_{\max}$, the alpha complex is the Delaunay triangulation.  The \emph{alpha shape} is the 
union of simplices in the alpha complex and for $\alpha \geq \alpha_{\max}$ the alpha shape is the 
convex hull of the point set. 

The alpha shape is a polytope in a fairly general sense: it can be concave and even disconnected.
Its components can be three-dimensional clumps of tetrahedra, two-dimensional patches of triangles, 
one-dimensional strings of edges, and collections of isolated points, as well as combinations of these 
four types. 

\begin{figure}
\centering
\includegraphics[width=3.25in]{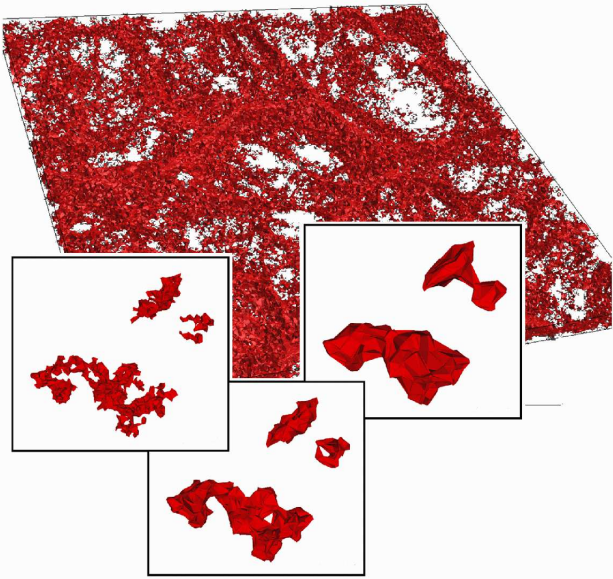}  
\caption{Top: $\alpha$ shape of a GIF $\Lambda$CDM simulation, slice through simulation box. 
Bottom: $\alpha$ shape at 3 different values of $\alpha$.}  
\label{fig:fig_ALPHA_2}
\end{figure}

Alpha shapes reflect the topological structure of a point distribution on a scale 
parameterized by the real number $\alpha$. Figure \ref{fig:fig_ALPHA_2} shows the $\alpha$-shape 
applied to a slice of a cosmological model. The three inserts show $\alpha$-shapes at three  
different $\alpha$ values.

\vspace{1.0truecm}
\subsection{Filtrations}
The concept of ``filtration'' is an important source of information about
the topology of a point distribution.  A filtration provides a view of the
topology as a function of scale.  Formally, 
given a space $\Manifold$, a \emph{filtration} is a nested sequence of subspaces:
\begin{equation}
  \emptyset=\Manifold_0\subseteq\Manifold_1\subseteq\ldots\subseteq\Manifold_m=\Manifold .
\end{equation}
The nature of the filtrations depends, amongst other things, on the representation of the mass distribution. 
For a discrete point distribution, the set of \emph{alpha shapes}, ordered by scale, constitute a filtration 
of the Delaunay tessellation.



The set of real numbers $\alpha$ leads to a family of shapes capturing the intuitive notion 
of the overall versus fine shape of a point set. Starting from the convex hull gradually decreasing $\alpha$, 
the shape of the point set gradually shrinks and starts to develop enclosed voids. These voids may join to 
form tunnels and larger voids. For negative $\alpha$, the alpha shape is empty. 

Although the alpha shape is defined for all real numbers $\alpha$, there are only a finite number of different 
alpha shapes for any finite point set. In other words, the alpha shape process 
proceeds discretely with increasing $\alpha$, marked by the addition of new Delaunay simplices once $\alpha$ exceeds the 
corresponding threshold. 

\subsection{Computing Betti Numbers}

The standard algorithm in 2-D and 3-D for calculating Betti numbers of an $\alpha$-complex is due to 
\cite{delfinado1993}.  
The algorithm is efficient in the sense that only one traversal of the simplices making up the $\alpha$ complex is 
required. First, the Delaunay tessellation of the point set is constructed. 
Its simplices - vertices, edges, triangles - can be ordered  by the radius of the smallest sphere that touches 
the points defining the given simplex and contains no other data points.  From this ordering a nested 
sequence of subcomplexes can be constructed, each member of the sequence being constructed from the preceding 
one by the addition of a new simplex having a larger scale.  If for example the added simplex is an edge, it 
will either connect two previously disjoint components or close a loop (thereby creating a tunnel).

Cycling through all its simplices, in three dimensions the calculation is based on the following 
straightforward considerations. When a vertex is added to the alpha complex, a new component is created 
and $\Betti_0$ is increased by $1$. Similarly, if an edge is added, it connects two vertices, which 
either belong to the same or to different components of the current complex. In the former case, the edge 
creates a new tunnel, so $\Betti_1$ is increased by $1$. In the latter case, two components get connected 
into one, so $\Betti_0$ is decreased by $1$. If a triangle is added, it either completes a void or it 
closes a tunnel. In the former case, $\Betti_2$ is increased by $1$, and in the latter case, $\Betti_1$ 
is decreased by $1$. Finally, when a tetrahedron is added, a void is filled, so $\Betti_2$ is lowered by $1$. 
Following this procedure, the algorithm has to include a technique for determining
whether a $p$-simplex belongs to a $p$-cycle. For vertices and tetrahedra, this is rather trivial.
For edges and triangles, a more elaborate procedure is necessary \citep[][]{delfinado1993}.

For software that implements these algorithmic ideas, we resort to the Computational Geometry Algorithms 
Library, \cgal\footnote{\cgal is a \texttt{C++} library of algorithms and data structures for computational geometry,
  see \url{www.cgal.org}.} \citep{caroli2011,da2011}. 

\begin{figure}[h]
\vskip 0.5truecm
\centering
\mbox{\hskip -0.5truecm\includegraphics[width=3.45in]{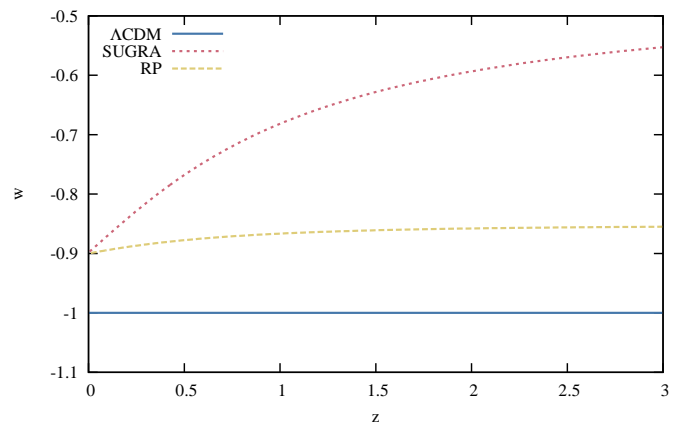}}
\label{fig:wzde}
\caption{Evolution of the dark energy equation of state parameter $w(z)$ for 
$\Lambda$CDM (solid), SUGRA (dotted) and RP (dashed).}
\label{fig:wz}
\end{figure}

\section{Topological Analysis of simulations}
\subsection{Our simulations}
We investigate the evolved structure in the standard $\Lambda$CDM model, the Ratra-Peebles (RP) model, and a SUGRA 
model. All simulations start with the primordial perturbation amplitudes and phases. The imprint of dark 
energy on the mass distribution is expressed via the different expansion history of the universe as a result of the 
differently evolving equation of state parameter $w(z)$ (fig.~\ref{fig:wz}). This is determined by the evolution of the 
quintessence scalar field $\phi$, forced by a parameterized potential $V(\phi)$. The DE model parameters 
(table 1 in \cite{bos2011}) are chosen such that the models differ significantly from $\Lambda$CDM, while still in 
accordance with observations. For example, the slope of the quintessence potentials is set to  $\zeta_\mathrm{RP}=0.347$ 
and $\zeta_\mathrm{SUGRA}=2.259$. 

The models we consider are described by WMAP7 \citep{wmap7} best-fit value parameters ($h=0.704$, $\Omega_m=0.272$, 
$\Omega_\Lambda=0.728$, $\sigma_8=0.809$ and $n_s=0.963$). For each model we ran $512^3$ and $256^3$ particle 
simulations, in a periodic box of $200h^{-1} \mathrm{Mpc}$ width, starting at $z\approx63$. 
The baryonic content of the universe is incorporated as dark matter particles, using 
$\Omega_m=\Omega_{dm}+\Omega_b=0.272$. 

We use the SUBFIND algorithm \citep{springel2001} to identify the gravitationally bound haloes in the resulting 
particle distributions. 
At $z=0$ the simulation boxes contain 170740 ($\Lambda$CDM), 172949 (RP) and 175332 (SUGRA) halos with mass higher than 
$M \sim 1.4\times 10^{11} h^{-1}{\rm M}_{\odot}$ ($>32$ particles). They trace the general structures present in the field, even 
though a lot of the detailed structure is lost or diluted (fig.~\ref{fig:SIMULATIONS}). 

Slices from two simulations, at $z=0$, are shown in figure \ref{fig:SIMULATIONS}.  The left hand panel displays 
the standard $\Lambda$CDM model and the right hand panel shows the SUGRA model.  
The slices as shown in the upper panels are perceptually 
indistinguishable.  Differences can only be perceived at higher resolution and lower contrast: this is shown in the lower panels 
of figure~\ref{fig:SIMULATIONS}, for the dark matter density field as well as the haloes. Different filaments appear, the 
subclustering is different in the big cluster and there are a number of different voids.  The voids differ in size and shape, 
and their degree of merging \citep[cf.][]{shethwey2004}.

\begin{figure}
\vskip -1.1truecm
\centering
\mbox{\hskip -1.0truecm\includegraphics[width=3.45in]{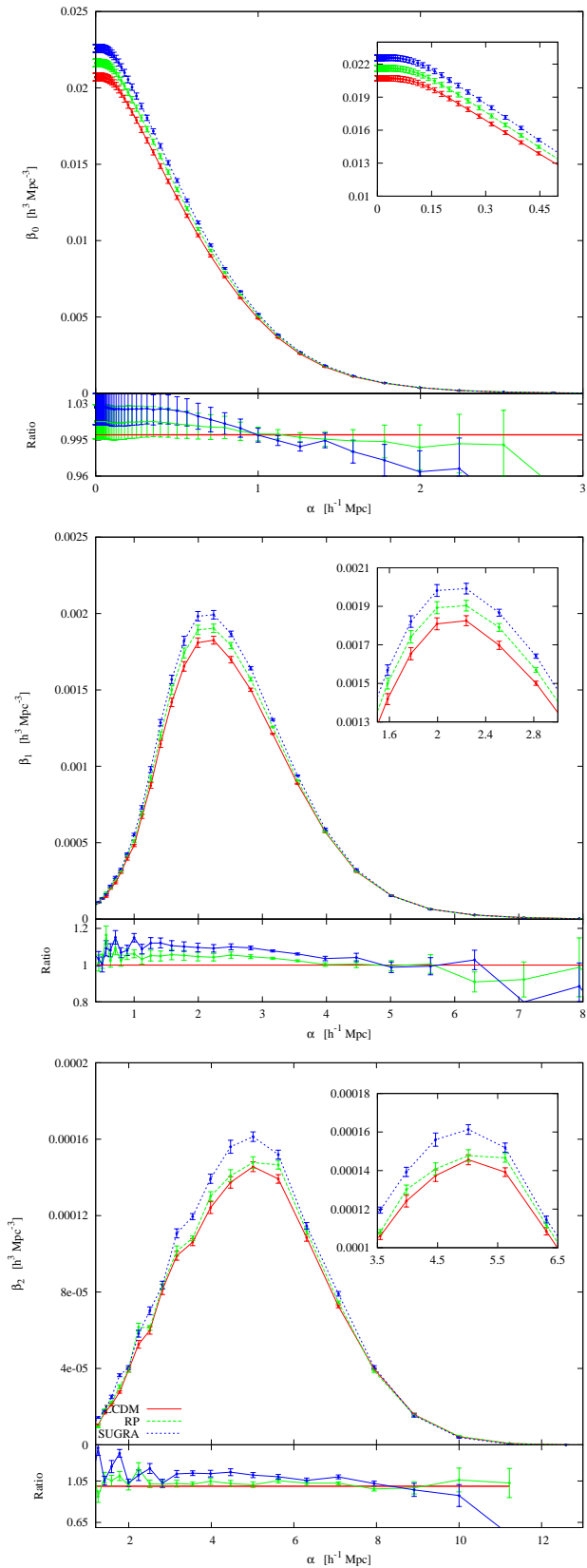}}
\label{fig:signatures}
\caption{Betti signatures: Betti numbers, per unit volume, as a function of scale $\alpha$ 
 at $z=0$. Red: $\Lambda$CDM, Green: RP, Blue: SUGRA. Top: $\beta_0$; Middle: $\beta_1$; Bottom: $\beta_2$. 
Inserts: detailed differences near distribution peak. Lower strip: $\beta_i / \beta_i(\Lambda$CDM).}
\vskip -0.5truecm
\end{figure}

\subsection{The analysis - Betti signatures}
We analyse the clustering of the haloes in these three simulations using \textit{Betti Numbers}. 
In order to compute the Betti numbers for a 
point distribution in 3 dimensions, we generate the Delaunay tessellation from the point sample and from this the 
nested sequence of {\it alpha shapes}. Subsequently, we determine the Betti numbers of the set of alpha shapes as a 
function of $\alpha$.  The resulting Betti number versus $\alpha$ curves are the \textit{Betti signature} of the point distribution. 
In Figure~4 we show the Betti signatures for the scale dependence of $\beta_0(\alpha)$, $\beta_1(\alpha)$, and 
$\beta_2(\alpha)$ in the three simulations. The error bars in all panels are derived on the basis of estimates from 8 sub-cubes.
There are significant differences between the curves for $\beta_1(\alpha)$, and $\beta_2(\alpha)$ in the three 
models, showing that the Betti signature is a sensitive discriminator of structure, even in a single redshift slice. This 
remains true when taking into account differences in $\beta_0$ and $\sigma_8$. In subsequent 
papers we follow the evolution of the Betti signature with time and assess their dependence on halo mass and show that it 
provides a sensitive measure of the nature of dark energy.

\section{Discussion and Conclusions}

We have presented a powerful topology-based method for classifying the structure that develops in cosmological 
N-body simulations.  The Betti signatures 
are an effective way of discriminating between models that superficially look the same. 
The method is an important generalisation of the genus and Minkowski functional approaches, in that Betti analysis of 
alpha shapes is able to discriminate structural content at a deeper level \citep{robins2006,weygaert2010,weygaert2011}. 


The Betti signatures of the cosmological simulations are shown to be capable of discriminating between models that are 
difficult to distinguish by other means. This may involve dark energy signatures or traces of primordial non-Gaussianities 
in the cosmic mass distribution. This is because dark energy affects the expansion of the universe, which affects the 
formation of nonlinear objects, while non-Gaussian initial conditions will change the formation history of galaxies and 
thus Betti numbers.

It is also important to put our approach in terms of Betti analysis in perspective relative to other measures of structure, 
in particular the Minkowski functionals \citep{mecke1994,schmalzing1997}.  While the Euler characteristic and Betti numbers 
give information about the connectivity of a manifold, the other three Minkowski functionals are sensitive to local manifold 
deformations.  The Minkowski functionals give information about the metric properties of a manifold on the basis of which 
broad-brush structural characteristics such as shape can be defined \citep{sahni1998}. The Betti numbers focus on its 
topological properties, the scale dependence likewise providing some metric information.  

In a series of following papers we shall be adding the powerful notion of {\it topological persistence}  
\citep{edelsbrunner2002} to our analysis of filtrations \citep{pranav2011}. Betti analysis is an integral 
part of an intrinsically much richer  topological language which addresses the hierarchical substructure 
of the Cosmic Web in an elegant and natural description \citep[for a cosmological application, see][]{sousbie2011a,
sousbie2011b}. In standard practice, the multiscale nature of the mass distribution tends to be investigated by means of 
user-imposed 
filtering. {\it Persistence} entails the conceptual framework and language for separating scales of a spatial structure, and 
rationalizes the multiscale approach by considering the range of filters at once. At the same time, it deepens the approach 
by combining it with topological measurements. Within the context of hierarchical cosmic structure formation,
persistence therefore provides a natural formalism for a multiscale topology study of the Cosmic Web. Alpha shapes provide 
the perfect context for understanding the concept: the position of a feature within the structural hierarchy is determined by 
the $\alpha$ interval over which it persists. Persistence even provides a natural path towards removing 
topological noise, which would be identified as features with small persistence \citep[see][]{weygaert2011}. 

\acknowledgments
We thank Manuel Caroli for the CGAL based software for computing 
alpha shapes and Betti numbers. RvdW, PP, GV and MW gratefully acknowledge the hospitality of INRIA/Sophia-Antipolis and 
financial support by the OrbiCG INRIA Associate Teams program. Part of this project was carried out within the context of 
the CG Learning project. The project CG Learning acknowledges the financial support of the Future and Emerging Technologies 
(FET) programme within the Seventh Framework Programme for Research of the European Commission, under FET-Open grant number: 
255827.

\end{document}